\documentclass[preprint,
             preprintnumbers,12pt,eqsecnum,
           tightenlines,a4paper,
           nofootinbib]{revtex4}
\usepackage{graphicx}

\newcommand{\beq}{\begin{equation}}
\newcommand{\eeq}{\end{equation}}
\newcommand{\bea}{\begin{eqnarray}}
\newcommand{\eea}{\end{eqnarray}}

\begin{document}

\
\vspace{.3cm}
\preprint{KYUSHU-HET-68}


\title{\LARGE Effective de Sitter brane\\ via running radius }

\author{
Nobuhiro Uekusa}
\email[]{uekusa@higgs.phys.kyushu-u.ac.jp}
\affiliation{
Department of Physics, Kyushu University, Hakozaki, 
Higashi-ku, Fukuoka 812-8581, Japan\\ \ \\  }



\begin{abstract}

\vspace{.3cm}

We present the solution for effective couplings on a 
running de Sitter brane.
A renormalization group formalism 
is developed for linearized gravity in a two de Sitter brane model.
It is shown that the effective tension 
remains a constant similarly
to the zero cosmological constant case.
We also discuss the suppression of higher derivative terms.


\end{abstract}


\maketitle 


\section{Introduction}

The holographic renormalization group 
plays an important role 
to gain a deep understanding of effective field theory.
This is motivated by the AdS/CFT conjecture
where
the size of an extra dimension is identified with the renormalization 
group scale in dual field theory \cite{Mal,GKP,Wit}. 
In particular, 
it becomes a useful tool when the calculation of field theory
is complicated.
Much attention has been attracted by the applications to
phenomenologically interesting brane world context 
such as the Randall-Sundrum (RS) model \cite{RS,Ver,BK,DL}.

A renormalization group formalism has been developed
for theories which have a fixed position of visible brane  
but a variable position of  hidden brane \cite{LMS,LR}.
It simplifies the calculation of
running couplings for localized operators  
generated by quantum corrections 
\cite{GGH,DGP}.
These analyses were applied to the case
where cosmological constants on branes are zero. 
Since the observation favors a positive cosmological constant \cite{WMAP}
and 4d Newton potential 
on de Sitter (dS) branes has been confirmed \cite{GNY},
we are interested in the case where 
cosmological constants on branes are nonzero.
The purpose of the present paper is to apply 
the renormalization group formalism 
to a model with two dS branes.

\vspace{.3cm}
In this paper, 
we vary the distance 
between two dS branes located at fixed points of 
the orbifold S$^1$/Z$_2$, 
while we keep the same physics on the visible brane.
The requirement that bulk classical field equations do not change
leads to 
the variation of the coefficients of hidden brane operators.
We give the exact solution for the 
coefficients 
of localized quadratic graviton
using a derivative expansion, 
where the corresponding
differential equation has the smooth zero cosmological constant limit.
In particular, 
we find out the radius independent tension of the effective hidden brane. 
This is similar to the case of 
the RS branes \cite{LR}. 
We also confirm that  higher derivative terms 
in the effective hidden brane action are suppressed 
compared to a localized Einstein-Hilbert term 
against a small variation of the radius.

The paper is organized as follows.
In Section \ref{s2}, we present the solution for couplings
on effective dS brane.
In Section \ref{s3}, we investigate the suppression of higher derivative terms.
Summary and discussions are given in the final section.

\section{\label{s2}Couplings on effective de Sitter brane}

We begin with the gravitational action,
\bea
  &&\qquad 
      S=S_{\textrm{\scriptsize bulk}}
        +S_{\textrm{\scriptsize hid}}
        +S_{\textrm{\scriptsize vis}} \ ,\label{act}
\\
  &&   S_{\textrm{\scriptsize bulk}}={1\over 2\kappa^2}\int d^4x dy\sqrt{-g} 
      (R-2\Lambda)  \ ,
\\
  &&\quad
    S_{\textrm{\scriptsize hid}}=
      -\tau_h\int d^4x \sqrt{-g_h}  \ ,    \label{hid}
\\
  &&\quad  
    S_{\textrm{\scriptsize vis}}= -\tau_v\int d^4x \sqrt{-g_v} \ ,
\eea
where $\Lambda(<0)$, $\kappa$, $\tau_h$ and $\tau_v$ denote
the bulk cosmological constant, 
the five-dimensional gravitational coupling constant,
the tensions of the hidden and visible branes, respectively.
The hidden and visible branes are located at $y=y_h$ and $y=y_v$
which correspond to two fixed points of the orbifold 
S$^1$/Z$_2$.
The determinants of the metrics in the  
bulk and on the hidden and visible branes
are denoted as  
$g$, $g_h$ and $g_v$, respectively.

For the action (\ref{act}),
the background solution with positive cosmological constants on
the hidden and visible branes is given by\footnote{The 
localization of various bulk fields
on a brane with a positive cosmological constant
has been investigated for 
the background (\ref{ba}) \cite{GNY,BGO,ue}.}
\beq
    ds^2=A^2(y)(-dt^2+a^2(t)\delta_{ij}dx^i dx^j)+dy^2 \ ,\label{ba}
\eeq
\beq
     A(y)={\sqrt{\lambda}\over\mu}\sinh\left[\mu(y_H-|y|)\right] \ , \label{wa}
\eeq
where $\mu=\sqrt{-\Lambda/6}$. 
The constant of integration $y_H$ is determined by 
the boundary condition $A(y_h)=1$.
The feature of the background (\ref{ba}) is summerized as follows.
The cosmological constants on the visible and hidden branes are 
written equally as
\beq
    \lambda=\left({\dot{a}\over a}\right)^2 \ .  \label{la4}
\eeq
The terminology ``dS brane ''
means a brane with positive cosmological constant $\lambda$.
As a consistency condition,
the cosmological constants on these branes
are related to
the parameters in the action (\ref{act}),
\beq
    \lambda={\Lambda\over 6}+{\kappa^4\tau_v^2\over 36} \ .
\eeq
In addition, the brane tensions are constrained as
\beq
    \tau_h=-\tau_v \ .
\eeq
As seen from the warp factor (\ref{wa}),
the location at $y=y_H$ is interpreted as a horizon. 
In the following, the region of the fifth dimension is 
taken as $y_h\leq y \leq y_v<y_H$.

\vspace{.5cm}

We consider the gravitational fluctuation around the background (\ref{ba}),
\beq
    ds^2=A^2(y)(-dt^2
    +a^2(t)\left[\delta_{ij}+h_{ij}(x^{\mu},y)\right]dx^i dx^j)+dy^2 \ ,
                                              \label{flu}
\eeq
where $x^{\mu}=(t,x^i)$.
This is convenient for the present calculations although
in the flat case, the fluctuation is often written as 
$A^2\eta_{\mu\nu}+h_{\mu\nu}$ instead of
$A^2(\eta_{\mu\nu}+h_{\mu\nu})$.
Under the gauge conditions, 
$h_i^i=0$ and $\partial_i h^{ij}=0$,
the field equation derived from the equations (\ref{act}) and (\ref{flu}) 
is given
by
\beq 
    \nabla_M^2 h_{ij}   =0 \ ,
                                           \label{buleq}
\eeq
where $\nabla_M$ denotes the five-dimensional covariant derivative.
In the equation (\ref{buleq}), we work with a system of 
only gravity
and the distance between the two branes is not dynamical.
We assume the proper radius is chosen 
so as to fit a possible explanation of the hierarchy  
as in the RS scenario \cite{RS}.
Thus the fluctuation in the bulk is completely determined by 
the equation (\ref{buleq}).

\vspace{.5cm}

According to \cite{LMS,LR},
we shorten the distance between the two branes
without varying the bulk field equation of the equation (\ref{buleq})
with fixed $y_v$.
Then effective local terms can be induced on the running hidden brane.
We would like to derive the radius dependence of
these local terms. 
To investigate it,
we introduce the effective hidden brane $S_l$ located at $y=l$ $(\geq y_h)$. 
In other words, 
by replacing $S_{\textrm{\scriptsize hid}}$
 in the action (\ref{act})
with $S_l$,
we obtain the effective theory in $l\leq y\leq y_v$,
\beq
   S_{\textrm{\scriptsize eff}}
    =S_{\textrm{\scriptsize bulk}}
     +S_l 
     +S_{\textrm{\scriptsize vis}} \ .   \label{eff}
\eeq
As the fluctuation $h_{ij}$ is small,
the quadratic action
of the effective hidden brane at $y=l$ is given
by  
\beq 
    S_l=-\tau_h\int d^4x
    \,{1\over 4}\, \alpha(x,l)\sqrt{-g_l}\,(A^2h)^2 \ , \label{aact}
\eeq
where 
$g_l$ denotes the determinant of the metric on the effective hidden 
brane.
The $\alpha(x,l)$ stands for a series of couplings multiplied
by derivative operators. 
In four-dimensional Fourier transforms 
$\alpha(x,l)\rightarrow \alpha(q,l)$, it is represented as 
\beq
   \alpha(q,l)=\sum_{j=0,1,\cdots} \alpha^{(2j)}(l)(q/A)^{2j} \ , \label{der}
\eeq
where momentum $q$ is defined as a small dimensionless quantity
divided by some large mass scale.
In the equations (\ref{aact}) and (\ref{der}), 
we omitted the explicit tensor structure and $a(t)$-dependence
since we are particularly interested in the position dependence of the coupling
$\alpha$.

\vspace{.3cm}

The linearized field equation derived from
the effective action $S_{\textrm{\scriptsize eff}}$ is given by\footnote{
In the flat case,
the factor $(3\alpha-4)$ of the equation (3.7) in \cite{LR}
can be found in the equation for the rescaled fluctuation $\hat{h}=A^2 h$,
where
$\partial_y^2 h=4\mu\delta(y-l)\hat{h}/A^2+(\textrm{no delta functions})$
and $\kappa^2\tau_h=6\mu$ are used.
}
\beq
     \left(A^{-2}\nabla_{\mu}^2
    +\partial_y^2+4{(\partial_y A)\over A}\partial_y
    -2\kappa^2\tau_h (\alpha-1)\delta(y-l)
   \right)h =0  \ .\label{fieq}
\eeq
From the equation (\ref{fieq}), the boundary condition at $y=l$ is obtained as 
\beq
      \partial_y h(y)|_{y=l}=\kappa^2 \tau_h (\alpha-1) h(l) \ . 
\eeq
Combining this with the field equation (\ref{fieq})
leads to the 
differential equation for $\alpha$,
\beq 
    {\partial\over \partial l}\alpha(l)=-\kappa^2\tau_h(\alpha-1)^2
        -4{A'\over A}(\alpha-1)+{q^2\over \kappa^2 \tau_h A^2} \ .  \label{flowal}
\eeq
where $A=A(l)$ and ${}'=\partial/\partial l$.
When we take the $\lambda\rightarrow 0$ limit 
$A=e^{-\mu l}\equiv 1/(\mu z)$, the equation (\ref{flowal})
reduces to
\beq
   z{\partial\over \partial z}\alpha
   =-6(\alpha-1)^2+4(\alpha-1)+{(qz)^2\over 6} \ ,
\eeq
where we used $6\mu=\kappa^2\tau_h$.
This agrees with the equation in the flat case derived in \cite{LR}.

\begin{figure}[htbp]
\begin{center}
\voffset=15cm
  \includegraphics[width=11cm,height=8cm]{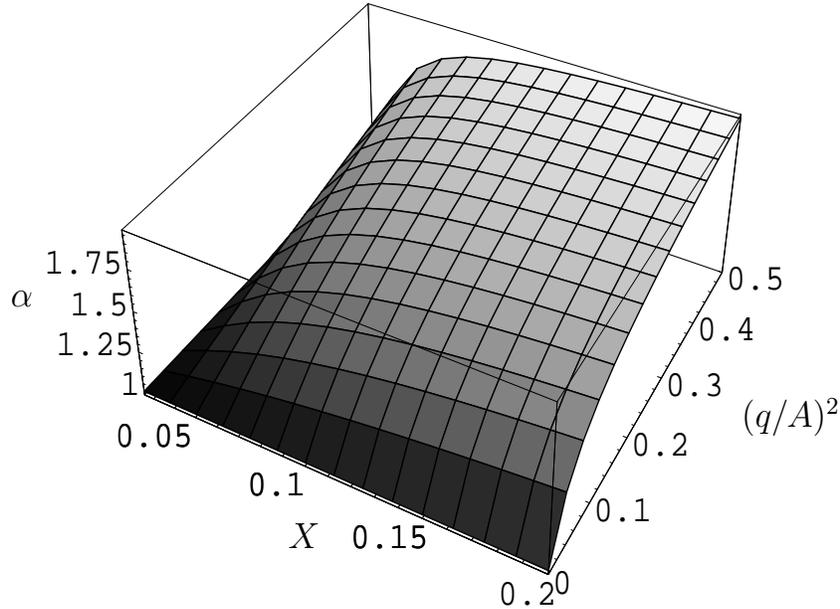}
\caption{$X$- and $(q/A)^2$-dependences of the coupling $\alpha$,
where
$\kappa^2\tau_h=1.2$, 
$\mu=\sqrt{390}\times 10^{-2}$, $\lambda=10^{-3}$ and $X_h=1/39$.
\label{alq}}
\end{center}
\end{figure}

We can find the exact solution for 
the equation (\ref{flowal}),
\beq
    \alpha=1+{2\mu\over\kappa^2\tau_h} X\sqrt{1+X}
 \left[
   {2\over X}-
    {{\beta_{+}\beta_{-} \over 3}
         F(1+\beta_{+},1+\beta_{-},4;-X)
       -B G_{22}^{20}\left(-X \left|
       {\begin{array}{cc}
        {}_{-\beta_{+}} & {}_{-\beta_{-}}\\
        {}_{-3}&  {}_{0}   \\
      \end{array}}\right.\right)
      \over 
      F(\beta_{+},\beta_{-},3;-X)
      +B G_{22}^{20}\left(-X \left|
         \begin{array}{cc}
        {}_{1-\beta_{+}} & {}_{1-\beta_{-}}\\
       {}_{-2}&  {}_{0}   \\
      \end{array}\right.\right)}\right] \ , \label{alpha}
\eeq
\beq
    X={1\over \sinh^2\left[\mu(y_H-l)\right]} \ ,\qquad
  \beta_{\pm}={1\over 4} \left(5\pm 3\sqrt{1+{4q^2\over 9\lambda}}\right) \ ,
\eeq
where
we introduced the new postion coordinate $X$
such that smaller $X$ is closer to the original hidden brane,
while $X\rightarrow \infty$ corresponds to the horizon.
The functions
$F(\delta,\beta,\gamma;-X)$ and 
$G_{22}^{20}\left(-X \left|
         \begin{array}{cc}
        {}_{\delta} & {}_{\beta}\\
        {}_{\gamma}&   {}_{0}   \\
      \end{array}\right.\right)$
stand for the Gauss's hypergeometric function and 
Meijer's $G$-function, respectively.
The constant of integration $B$ is determined by 
the boundary condition $\alpha(y_h)=1$ from the equation (\ref{buleq}) 
and (\ref{fieq}),
\beq
   B=-{
      F(\beta_{+},\beta_{-},3;-{\lambda\over \mu^2})
    -{\beta_{+}\beta_{-}\over 3}{\lambda\over 2\mu^2}
       F(1+\beta_{+},1+\beta_{-},4;-{\lambda\over \mu^2})
    \over
          G_{22}^{20}\left(-{\lambda\over \mu^2}\left|
           \begin{array}{cc}
        {}_{1-\beta_{+}}& {}_{1-\beta_{-}} \\
         {}_{-2}& {}_{0}\\
        \end{array}
      \right.\right)
    +{\lambda\over 2\mu^2}G_{22}^{20}\left(-{\lambda\over \mu^2}\left|
         \begin{array}{cc}
         {}_{-\beta_{+}}& {}_{-\beta_{-}} \\
         {}_{-3}  & {}_{0}\\
        \end{array}
      \right.\right)} \ .
\eeq
Fig.\ref{alq} shows
the behavior of our solution (\ref{alpha}) for the coupling 
$\alpha$
against the position $X$ as well as the derivative operator $(q/A)^2$. 
The coupling $\alpha|_{q^2=0}$
is independent of
the position and the slope
$\partial_{(q^2/A^2)}\alpha|_{q^2=0}$ is nonzero
except at $X(l)=X(y_h)\equiv X_h$.
This means that the tension of the effective hidden brane,
$\tau_h\alpha^{(0)}$, is 
constant and
that the localized curvature term appears on the effective hidden brane.
The radius independent behavior of the effective hidden brane tension
is 
similar to
the case of RS branes. 

\vspace{.3cm}

It is useful to take some limits  
in order to study properties of the solution (\ref{alpha}).
We look at the coupling $\alpha$ in 
two limits, $q^2=0$ and large $X$.

For $q^2=0$, 
the equation (\ref{flowal}) reduces to
\beq 
    {\partial\over \partial l}\alpha^{(0)}(l)
   =-\kappa^2\tau_h(\alpha^{(0)}-1)^2
        -4{A'\over A}(\alpha^{(0)}-1)
  \ .  
\eeq
The general solution for this equation is given by
\beq
   \alpha^{(0)}=1+
    {3\mu\over \kappa^2\tau_h}
   {X^2\over (X-2)\sqrt{1+X}+C} \ ,
\eeq
where $C$ is a constant of integration.
By imposing the boundary condition 
$\alpha^{(0)}(y_h)=1$, we obtain the solution,
\beq
   \alpha^{(0)}=1 \ .
\eeq
Thus the value of the hidden brane tension $\tau_h$
is preserved in the variation of the radius. 

\vspace{.3cm}
For a large $X$ limit,  
the coupling $\alpha$ is given by
\beq 
    \alpha \simeq {4\mu\over \kappa^2\tau_h}\sqrt{X}
       (1-{\cal F}(q^2)) \ ,  \label{alx}
\eeq
where ${\cal F}(q^2)$ denotes the coefficients dependent on $q^2$.
Since ${\cal F}$ depends on $q^2$ only through the form
$\sqrt{1+4q^2/(9\lambda)}=\sqrt{1+4q^2/(9\mu^2X A^2)}$,
we can identify   
a ratio of higher derivative terms,
\beq  
    \alpha^{(2j+2)}\left({q\over A}\right)^{2j+2}\left/
    \alpha^{(2j)}\left({q\over A}\right)^{2j}     \right.
    \propto
     -{q^2\over \mu^2 X A^2}=-{q^2\over \lambda}\ ,\qquad\textrm{for } j>0\ .
\eeq 
This result seems to imply 
that for tiny $\lambda$, the contributions of
higher derivative terms in
the action (\ref{aact})
can exceed the Einstein-Hilbert term (i.e. the $q^2$ term)
far away from the original position of the hidden brane.
However before
concluding the breakdown of linearized approximation,
one must require to take KK-modes into account 
since 
KK-modes are localized 
near 
the visible brane
\cite{GKR}.  
We leave this problem in future.
In next section, we will study the contributions
of higher derivative terms  at $l\sim y_h$.

\section{\label{s3}The suppression of higher derivative terms}

We would like to examine the radius dependence of 
$\alpha$
at each order of derivative expansion.
We need to check particularly
whether the contributions of higher derivative terms are negligible. 
Since it is difficult to calculate
directly a derivative expansion of our solution (\ref{alpha}),
we return to the 
differential
equation (\ref{flowal}).

%

Expanding the 
equation (\ref{flowal}) about $(q/A)^2$,
we obtain a set of the 
equations as
\bea
   &&\alpha^{(2)}{}'+2{A'\over A}\alpha^{(2)}+
      -{1\over \kappa^2\tau_h}=0 \ ,
\nonumber\\
   &&\alpha^{(4)}{}'
     +\kappa^2\tau_h \alpha^{(2)}{}^2
    =0 \ ,
\nonumber\\
   && \cdots \ ,
\nonumber\\     \label{set}
\eea
where 
the ellipsis corresponds to
the equations for $(q/A)^{2n}$, $n>2$.
The 
equations in 
(\ref{set}) are numerically solved.  
%
\begin{figure}[htbp]
\begin{center}
\voffset=15cm
  \includegraphics[width=8cm,height=6cm]{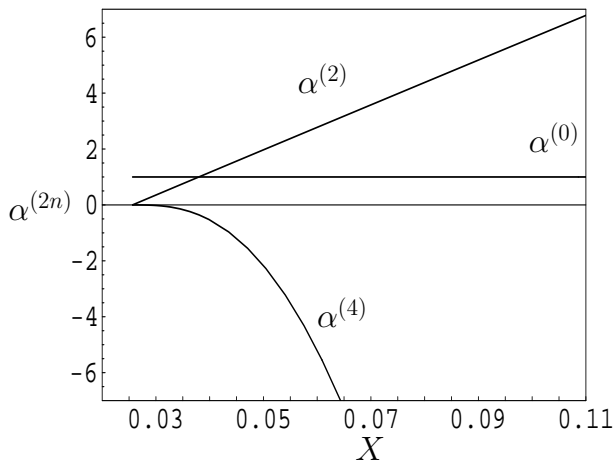}
\caption{$\alpha^{(2n)}$-$X$ graph for $n\leq 2$,
where
$\kappa^2\tau_h=1.2$, $\mu=\sqrt{390}\times 10^{-2}$, $\lambda=10^{-3}$
and $X_h=1/39$.
\label{alqq}}
\end{center}
\end{figure}
%
The position dependences of the couplings
$\alpha^{(0)}$, $\alpha^{(2)}$ and $\alpha^{(4)}$ 
are shown in Fig.\ref{alqq}.
For small $X$,
higher derivative terms are 
of the order of lower derivative terms
up to ${\cal O}((q/A)^2)$.
Therefore as long as we consider 
a small variation of the radius,
where $A\sim 1$ is satisfied,
the contributions of higher derivative terms are suppressed by 
powers of $q^2$.

\section{Summary and discussions}

In this paper, we have found the solution for  
effective couplings
on the running dS brane for linearized gravity.
The corresponding differential equation has
the smooth $\lambda\rightarrow 0$ limit.
The solution 
is a constant
at zero-th order 
(corresponding to the effective tension)
of a derivative expansion 
similarly
to the case of a RS brane model. 
In addition, 
an Einstein-Hilbert term and higher derivative terms appear
in the effective hidden brane action with radius dependence.
We have 
confirmed up to ${\cal O}(q^6)$ 
that higher derivative terms 
are suppressed against a small variation of the radius.
Far away from the original position of the hidden brane,
the KK-mode contribution becomes significant.
%
%
%
As for KK-modes,
it would be important to investigate not only the contributions
of higher derivative terms
but  
the spectrum of the KK gravitons 
which might be modified near the visible brane \cite{DHR}.


\vspace{.3cm}

We finally note that
the bulk background in our model is not AdS space 
in order to construct dS branes.
Beyond the AdS/CFT conjecture,
it remains to be examined whether some dual field theory exists.
This would require explicit calculations of quantum corrections
in field theory.
If dual theory  exists, 
it is worthwhile to examine whether and how conformal symmetry is broken. 

\section*{Acknowledgement}

The author thanks K.~Yoshioka and K.~Harada for reading the manuscript and discussions.
He is also grateful to K.~Ghoroku and K.~Inoue for discussions.


\end{document}